# A STUDY ON PRE-SERVICE PHYSICS TEACHERS' CONCEPTUALIZATION ON ELEMENTARY QUANTUM MECHANICS

*Um estudo sobre a conceitualização de futuros professores de Física em Mecânica Quântica elementar*

**Glauco Cohen Ferreira Pantoja** [glaucopantoja@hotmail.com]
*Instituto de Ciências da Educação
Universidade Federal do Oeste do Pará
Avenida Marechal Rondon, S/N, Pará, Brasil*

**Abstract**

In this work, we present the results of a research in which we aimed to evidence obstacles and advances in pre-service teachers' conceptualization on a subject involving elementary Quantum Mechanics. We based our analysis on the theories due to David Ausubel and Gèrard Vergnaud to study Meaningful Learning patterns, both in predicative and operatory form of knowledge, of six students involved in a didactical intervention composed of six classes, in which we emphasized both similarities and differences between Classical and Quantum Physics. With this intervention, we intended to teach the concepts of Physical System, Dynamical Variables, State of a Physical System and Time Evolution. We guided our data analysis by the methodology of content analysis (Bardin, 2008) and it turned possible to map Meaningful Learning patterns involving the four concepts to which were associated a set of essential features (in the predicative stage) and a set of theorems-in-action (in the operatory stage) relating the aim-concepts in problem-solving or conceptual mapping.

**Resumo**

Neste trabalho, apresentamos os resultados de uma pesquisa na qual focou-se evidenciar obstáculos e avanços na conceitualização de futuros professores em uma disciplina envolvendo Mecânica Quântica elementar. Baseou-se a análise nas teorias devidas a David Ausubel e Gèrard Vergnaud para estudar padrões de Aprendizagem Significativa, tanto na etapa predicativa como na etapa operatória do conhecimento, de seis estudantes envolvidos em uma intervenção didática composta de seis aulas nas quais enfatizou-se tanto similaridades e diferenças entre Física Clássica e Física Quântica. A intervenção abordou os conceitos de Sistema Físico, Variável Dinâmica, Estado de um Sistema Físico e Evolução Temporal. Guiou-se a análise de dados pela metodologia da Análise de Conteúdo (Bardin, 2008) e isto tornou possível mapear padrões de Aprendizagem Significativa envolvendo os quatro conceitos aos quais estavam associados um conjunto de atributos essenciais (na etapa predicativa) e um conjunto de teroremas-em-ação (na etapa operatória) relacionando os conceitos-foco em processos de resolução de problema ou de mapeamento conceitual.

**Palavras-chave:** Mecânica Quântica; Aprendizagem Significativa; Processos de conceituação.

## INTRODUCTION

Quantum Mechanics teaching is a subject that must be taken seriously if we really want to make our students aware of the achievements of modern science and if we wish to communicate the cultural scientific inheritance constructed, not only in the twentieth century, but that is still figuring as a progressive research program as Imre Lakatos would say. Research on Quantum Mechanics teaching, however, is not the mainstream of Physics Education Research (Çaliskan, 2009) and the curriculum doesn't emphasize the discussion of Modern Physics topics (Olsen, 2002). Although Greca and Moreira (2001) and Pantoja et al.

(2011) point a progressive development in this category of research, the authors stress this task hasn't been fulfilled yet.

Knowledge developed in Research on Quantum Mechanics Teaching, however incipient, must enlighten the path for researchers who want to understand teaching-learning processes better. One problematic feature in this context that seems to be consensual among researchers is the construction, due to students, of hybrid Mental Models juxtaposing quantum and classical characteristics. Tsarpalis and Papaphotis (2009), for example, state that Old Quantum Theory Models are distinct from those related to Modern Quantum Theory, classifying the former as desperate attempts of combining quantization of energy and classical mechanics pictures to give adequate answers to the crisis lived by Physics in the beginning of the twentieth century, saying it in a kuhnian language, while the latter are complete models used to microscopically describe[1] events or regularities occurring in nature.

Another feature that must be stressed is the excessive emphasis on mathematical formalization on Quantum Mechanics in undergraduate level. Some professors, based on what Mario Bunge (2011) calls operationalist view of science and of science education, do not approach Quantum Mechanics from its very fundamental concepts, but only from the formal (mathematical) point of view. They usually start defining the Quantum Mechanical State, for example, telling students it's a vector and emphasizing solely commutation relations instead of their physical content. They end up communicating an idea that Quantum Mechanics is nothing more than a set of linear algebra or differential equations problems, that is, traditional courses are all about developing mathematical calculations. Obviously, there's much mathematics underlying its' concepts, but there are possible rich discussions on ontology and epistemology of Modern Physics that are cast aside when the teaching process strongly relies only on mathematics. As it is stated by Greca (2000), this kind of approach is likely to lead some students to understand Quantum Physics formalism using their prior knowledge founded in classical mechanics, a relevant prior knowledge for learning physics, by the way.

Based on these fundamental ideas we tried to build a didactical unity to teach basic quantum concepts to students of Physics courses. In this paper, we discuss some possibilities and limitations in learning processes of six students enrolled in a subject on Modern Physics that emphasized a conceptual, rather than only mathematical, approach to Quantum Mechanics. This doesn't mean mathematics was eliminated, mainly because we tried to use as representation the Dirac notation commonly used in Quantum Mechanics undergraduate courses.

Our approach is basically guided by Meaningful Learning and Conceptual Field theories. Both emphasize the role of prior knowledge in conceptual development and learning but they tackle this problem in two complementary points of view. While Ausubel focuses knowledge in declarative form, Vergnaud studies the development of procedural or operatory knowledge. Both forms are explored in this work and we have based on their teaching-learning principles as discussed in the following theoretical framework section.

In this study, we present and comment data obtained in a research conducted with pre-service teachers. Both methodological and didactical approaches were similar to the ones by Pantoja et al. (2012) and to Pantoja et al. (2014), but with different students, even though these ones showed not much difference in prior knowledge compared to the ones involved in other studies. To analyze learners' relevant prior knowledge, the authors presented a questionnaire to the individuals in the research, and using the content analysis methodology, they could infer criterial attributes (Ausubel, 2000) attached to the prior relevant conceptual knowledge students brought.

**THEORETICAL FRAMEWORK**

Before carrying on the discussion on the Meaningful Learning Theory, we shall stress a subtle difference between two complementary forms of knowledge, one explicit, verbalized and formalized, and the other one, largely implicit, not so formalized but carrying conceptual knowledge with it. The former form of knowledge is known as predicative and the latter one is known as operatory.

The mentioned separation is important from an epistemological point of view: explicit knowledge can be negotiated, but the implicit one doesn't, once it underlies action. Conceptual Fields Theory, however, attributes Meaning to both forms of knowledge and, particularly, Vergnaud brings the idea that specific contents are also developed over time, as long as a person faces and then resolves diverse situations, in the operatory form. The development of operatory knowledge and of its' content is barely emphasized in literature, however Vergnaud has been evidencing, since a long time, the separation between procedural knowledge and specific content is unconceivable.

---

[1] Our reference here is to Quantum non-relativistic Mechanics, but it also can be applied to Field Quantum theory, Solid State Physics, Nuclear Physics, Atomic Physics, Relativistic Quantum Mechanics.

As we split these two forms of knowledge in two interdependent expressions of conceptualization, we work with the idea that knowledge is largely tacit and develops throughout time. We must then turn our attention to the mastering of situations, representations and Meanings associated to an organized body of knowledge. This is what Vergnaud calls a Conceptual Field, a set of situations, the mastering of which requires concepts, operations and representations. This is the reason why we do not approach the problem of acquisition of knowledge in Quantum Mechanics solely from an ausubelian point of view.

**Meaningful Learning theory**

We will discuss Meaningful Learning theory briefly, because has been extensively discussed in literature. We can see it in Ausubel (2000), Moreira (1997; 2003), Pantoja (2011), Rocha (2008), for example. We present the aspects we think that play the most important role in the problem we are working on. These ideas were crucial to the planning and implementation of the instruction and to the data analysis.

Ausubel conceives Meaningful Learning as being a process of acquisition of new knowledge in which there is the construction of Meaning. For him, it is an idiosyncratic psychological clear, conscious and differentiated entity. Ausubel classifies Meaning as mainly predicative. For the occurrence of Meaningful Learning, it's necessary both relevant prior knowledge in Cognitive Structure and a Potentially Meaningful Learning Material.

Ausubel postulates learning occurs in three stages. The first of them is acquisition, when new content is processed in someone's Cognitive Structure. This process is followed by retention, the storage of this information in such Cognitive Structure. Finally, the stage of obliteration occurs and then the "forgetting" of the subsumed information[2] happens. Meaningful Learning, which is generally associated to great rates of retention and low ones of obliteration, depends on two other idiosyncratic factors: the clarity prior relevant knowledge in the Cognitive Structure and the will of relating the new information to the old one in a non-arbitrary and substantive form.

Sometimes students have huge conceptual hindrances with their prior knowledge[3]. It can be unstructured, unclear, vague or poor and these aspects hinder Meaningful Learning, because it turns out to be difficult to anchor new knowledge to the Cognitive Structure. Once Meaningful Learning is constrained to clearness and specificity of the Subsumers, students like these may face great difficulties in initial learning, and sometimes are driven to rote learning processes.

Let us now discuss the learning processes in reception learning. Ausubel states that there are three kinds of elementary learning tasks[4]: representation learning tasks – acquisition and retention of entities that present objects in their absence; concept learning tasks – acquisition and retention of entities that can either have names or not, but are about regularities in events and objects; proposition learning tasks – acquisition and retention of combinations of words that have their own Meaning and express relations between objects. These three forms of learning are of great importance in physics teaching, once they are classes through which we can learn knowledge.

It is very clear that Cognitive Structures vary among people, and that's because knowledge can be achieved in a great number of forms. People experience different cultures, teachers, schools, books, and so forth. So, in one task you can see variable structures of knowledge presented; some of them, in a more explicit way and others in a more implicit manner, depending on the person who does it.

We acquire pieces of knowledge in a lifetime, but only part of them is relevant to a specific learning task. Ausubel calls Subsumers these relevant cognitive elements, and they compose part of students' prior ideas before any learning activity. In this view, you can't learn from what you don't know. These are distinct from each other, once they can be clear (precise), stable (coherent and wide), available (disposable for the realization of the cognitive task).

In the research process carried out in this study, we mainly analyzed, concerning predicative knowledge, Cognitive Structure variables, once they shed some light on the most difficult concepts. This exploration was of great worth once we could understand why some difficulties appeared and conjecture a way of "dismantle" them. Besides that, some gaps were to be filled and operatory knowledge could help understanding how apparently disconnected forms of reasoning are, in fact, attached.

---

[2] This process, in Meaningful Learning, is associated to residual information which is different of the initial one.
[3] Called Subsumers.
[4] By elementary we mean that all other predicative knowledge tasks can be reduced to these three forms.

To guide the instruction and manipulate the Cognitive Structure variables, Ausubel proposes four programmatic principles, which we have used in the designing of the didactic intervention. These principles are called *progressive differentiation*, stating the most general and abstract concepts must be presented first and, then, in order to elaborate or illustrate them with specific instances, less general concepts, propositions and representations are presented; *integrative reconciliation*, stating that the differences and similarities between apparently paradoxical instances must be stressed; *consolidation*, stating that the learning of a lesson must be made sure by using the same ideas in diverse contexts; and *sequential organization*, stating that which the content must be organized taking into account sequences of potentially Meaningful coherent units.

**Conceptual Fields Theory**

Now we shall discuss operatory knowledge, the one used in tacit manner. For Vergnaud, knowledge can be understood as a set of competences that develop progressively over time, that is, some of these competences evolve to a more complex form, while others are deteriorated. These competences are understood in terms of schemas, which are invariant organizations of actions or behaviors presented by someone when he or she faces a class of situations. Despite the feature of invariance, schemas are dynamical functional totalities and are not restrained to the sensory-motor activity, but also to the cognitive ones. So, grabbing an object and solving an electromagnetism problem may be classified both as actions in which are necessary schemas for its mastering.

It's important to note that a schema is not an invariant behavior, neither it's a stereotype. It is an invariant organization of actions, and that's because it is dynamical, once different actions can be organized in the same way. To count or to attribute functions to different people, for example, embraces the relation between pointing to an object and enumerating them.

Those schemas are composed basically of four elements:

- **goals and anticipations;**
- **action rules like "if x, then y";**
- **inferences;**
- **knowledge-in-action**

Goals and anticipations are conceived by Vergnaud as the expectations about the activity merged in the situations to be mastered. Subjects can describe some expected effects or phenomena (Vergnaud, 1996, p. 201). Both experts and novices can state or generate (implicitly) goals and anticipations for activities, but most of times experts will glimpse the "correct"[5] ones in shorter times than novices.

Action rules compose an important part of Schemas, once they allow the subject to continue actions that transform reality, storage of information and control the activity results. These processes play an important role in situation mastering, because they permit reaching the outcomes of activity in a context of permanent evolution.

Another substantial piece of Schemas is the possibility of inferences. They may allow subjects to "calculate" rules of action and anticipations from given information and available knowledge in the situation.

Prior knowledge, in the conception of Vergnaud, can be either explicit or largely implicit and is extremely content dependent. His emphasis is on the operatory form, but his account explains mastering of both forms. He divides knowledge into two basic complementary classes: propositions regarded by subjects as true about reality (theorems-in-action); categories and predicates, which are pertinent or not to conceptualize reality (concepts-in-action).

---

[5] By "correct" we mean scientifically accepted knowledge. No supposition on superiority of science is made. Science, for us, is a conceptual structure of reference, not an absolute truth.

There is an essential difference between concepts-in-action and theorems-in-action. To propositions, that is, theorems-in-action, we can attribute a truth value, while we can't do this with concepts, which are just pertinent or not. Concepts can be used to select important information by "filtering", through classification or grouping, important features of reality. On the other hand, the theorems-in-action are entities that establish important relations among classes and state truth or falsehood in relation to reality. Both are content dependent and, because of that, they are strongly biased by context[6].

For Vergnaud, the first moment of conceptualization comes from interactions among situations and schemas. It's warranted by the fact that mastering of a given situation is made by the "selection"[7] of schemas to progress in it. Relationship between the thinking subject and an arbitrary object whose properties must be abstracted is the dominant feature in the predicative form of knowledge. Insofar this level of conceptualization is the most formal and explicit of all, the ability of direct reference to an object, without considering specific instances, is achieved when people reach the most verbally advanced cognitive level and mastering of what Vergnaud calls a conceptual field.

For Vergnaud a conceptual field is

> *"…an informal and heterogeneous set of problems, situations, concepts, relationships, structures, contents, and operations of thought, connected to one another and likely to be interwoven during the process of acquisition." (Vergnaud, 1982, p.40).*

Vergnaud clarifies his ideas, introducing the idea of Concept, which is different of the idea of Concept-in-action. By concept he means a triplet of three sets

$$C = (S, K, R)$$

Where *S* is a set of situations that makes the concept useful and Meaningful, *K* is a set of knowledge-in-action that might be used by subjects when they face situations and *R* is a set of representations that may be linguistic, gestural, graphic and are used to represent S, K and other procedures necessary for solving S (Vergnaud, 1996, p.6).

Some representations are more powerful than others, but they can't be manipulated before the subject has grasped their Meaning. Graphics are, sometimes, and for some individuals, not so effective in representing as equations, though may be better understandable for some other people than the latter. It is desirable, however, that students could foster their capacity of using the greater number of representation as possible.

To put an end to this section we would like to stress again that subjects master a given conceptual field from interactions of his/her schemas with situations which he/she faces. Schemas, Vergnaud's main unity analysis, are composed of content dependent knowledge, that is, concepts-in-action and theorems-in-action, something that the piagetian theory doesn't regard, despite Vergnaud's consideration of the schemas as the best part of the Piaget's heritage.

**METHODOLOGY**

Now we shall make a detailed discussion concerning research and teaching methodologies. In the first part, we discuss the teaching methodology used and in the other one, we discuss research methodology used in data analysis.

**Teaching Methodology**

We designed and implemented a 12-hour didactic approach in a class with six undergraduate Physics students. All of them were studying in an introductory discipline on Modern Physics for the second time, once they took a course on a related subject in the prior semester. Students' prior knowledge seemed to be mostly related to solutions of Schroedinger's equation for distinct potentials.

---

[6] Here we do not refer to social context, but to the situation context. However, some investigation on this topic would lead to the same conclusion.
[7] By selection, we don't mean that it's deliberated. It can be even unconscious.

The content embraced by the proposal included the concepts of Physical Systems, Dynamical Variables, State of a Physical System and Time Evolution. We had two main reasons for that: 1) the theoretical principle of progressive differentiation, once these concepts are some of the most general in Quantum Mechanics and must be presented first (and then differentiated in a progressive manner - and after that, reconciled in a whole integrated structure); 2) the teaching-learning processes concerning the concepts of State of a Physical System and Time Evolution are rarely found in a bibliographic review (author et al., 2011). As exceptions we have, for example, in Singh (2008), Greca (2000), Rocha (2008), and author (2011).

The main goal of the instruction we've designed was the teaching of the concept of Time Evolution. Our reasons for choosing this so general concept are: 1) First, and scientifically speaking, this concept evokes two of great importance in physics, the ones of predictivity (the capacity of foreseeing effects) and causality (knowing the effects from the causes); 2) teaching on this topic, as said before, is rarely elected as a relevant research object, even in Physics Teaching Research. So, research on teaching processes regarding this concept is of great importance for Physics Education, in general, and for the Quantum Mechanics Teaching, specifically.

The concepts of State of a Physical System and of Dynamical Variables were considered as more general than the one of Time Evolution, once the latter is attached to the other two as a specific instance. We shall stress that both can vary along time and this two possibilities gave rise to Schroedinger's picture (State vectors change as time goes by) and Heisenberg's picture (Operators change as time passes by).

The didactical intervention started with the concept of Physical System, regarded as the most foundational of the four we've selected. Once it's a basic concept for classifying real objects, it's understood that in a proposal aiming to foster Meaningful Learning of Quantum Mechanics Concepts, it must be the first one to be taught, and by Quantum Mechanics it's regarded the Physical study of "tiny"[8] Quantum Physical Systems.

The intrinsic sequential organization of the concepts within the content has been considered. Thus, we discussed the following concepts (in the presented sequence): Physical System (the most general) → Dynamical Variables → State of a Physical System → Time Evolution. This sequence has facilitated, in our point of view, the development of the progressive differentiation and, lately, the integrative reconciliation processes.

The unity was also based upon the principles stated by Vergnaud that the mastering of a given conceptual field depends on the growing mastery of the situations which compose it. Therefore, we have chosen three classes of situations for the discussion: the Stern-Gerlach experiment, the hydrogen atom and the ammonia molecule. These situations have played an important role in both research[9] and teaching processes (Pantoja et al., 2012; Pantoja et al., 2014).

The reasons for choosing the Stern-Gerlach experiment are both pragmatic and epistemological. The practical one is linked to the conceptual two-level representation in Quantum Mechanics for spin 1/2 systems, the simplest ones (mathematically speaking) of sufficient richness for treating quantum mechanical features. With this kind of Physical System, it is possible to discuss the: Incompatibility or Compatibility of certain Dynamical Variables; Measurement Problem; State Superposition Principle; Angular Momentum in Quantum Mechanics; Notions of Spin Projections; and so on. As Vergnaud usually says, one Situation can give rise to distinct Concepts.

It is pointed in the literature (Kalkanis et al., 2002) that the Bohr's atomic model of the hydrogen atom is strongly bounded in student's Cognitive Structure, despite its' peculiar classical features. Considering it the prior knowledge of students on the structure (and casting aside its' dynamics, which we did not reinforce in the proposal) of the system, was of great value for this research. An epistemological reason for choosing this concept is the same of the Stern-Gerlach experiment, namely, the one of making sense of some important and central concepts.

---

[8] Physical objects embracing molecules, atoms, particles, whose characteristics (mass, electric charge, spin, radius, momentum) are relevant in the $\hbar$ limit. Empirically, they compose models giving precise results for problems whose solutions are not well explained by Classical Physics in the beginning of the twentieth century. We understand it's important to highlight this to cease ambiguity and circularity in the definition of Quantum Object as the one which obeys the laws of Quantum Physics.

[9] As we shall stress in the results, through the knowledge-in-action analysis.

The contribution of the ammonia molecule is foremost empirical. It can be used in laboratory to highlight the State Superposition Principle. We used it as an example of phenomenological context in theoretical approach to Quantum Mechanics. We could introduce some purely Quantum Mechanics aspects, for example, the quantum tunneling effect, when it was discussed.

We shall stress the dialectic feature between concept and situations on the proposal. In first place we presented the situation, the concepts that would be discussed, and finally, the processes of progressive differentiation and integrative reconciliation. Then, another situation was presented and the process was carried out the same way.

An important point to emphasize is the consideration on students' relevant prior knowledge. We supposed that they already knew something on classical processes of Time Evolution, a fact that was corroborated by the pre-test. This assessment showed a high level of generality for the mentioned concepts and the problem-solving analysis has suggested that students had more confidence in their Classical Physics knowledge (specially the one on Classical Mechanics) than in their Quantum Mechanical counterpart.

An instruction material was designed for the proposal and can be encountered in Pantoja (2011). Class planning included presentation of situations for the students to express their understanding of concepts that would be discussed in classroom after the teacher immerged the content in context. Then the processes of progressive differentiation and integrative reconciliation could be put in practice.

**Research Methodology**

We followed a research methodology with qualitative orientation, which is attached to a naturalist paradigm. This philosophy understands the world as having complex features that evidence a great number of forms (Bogdan and Biklen, 1994). We focused Meaning as the central variable in research and we considered that it is idiosyncratic, an aspect that focused the point of view of data analysis on the fact that people perceive the world in different forms, however there can be similarities along the process.

We adopted the triangulation methods to raise the level of credibility (an analogous entity of the one of reliability on quantitative research). This way, we believe we have lowered the subjectivity inherent to the qualitative data analysis. As research instruments, we have adopted pencil-and-paper problems, field note registers and conceptual map construction. We shall stress that we attributed fictitious names to each student to keep their identities in secret.

We have divided the data analysis in two large groups. The first to be discussed is the one in which we have studied the acquisition of predicative knowledge (verbal and organized). The other is the one in which we have studied both implicit thought operations and possible knowledge-in-action involved in problem-solving processes. Both analyses were guided by the content analysis methodology (Bardin, 2008).

The students answered to six paper-and-pencil problems and drew concept maps. Thus, we considered these activities as the corpus of the research. Further conclusions or evidences were obtained by our personal field notes. The inferred variables were considered as criterial/essential attributes (Ausubel, 1980) of the concepts in predicative knowledge analysis, while in operatory knowledge analysis, possible knowledge-in-action played the role of inferred variables.

The process of research was started with the pre-analysis, in which we made a "floating reading" consisting of relating information for inference making. After this step, we started the process of organizing data using index/indicators technique (Bardin, 2008). Finally, we've completed a cycle of research establishing the inferences presented here.

In the predicative knowledge analysis, our main goals were the inquiry on assimilation patterns, which means the product of non-arbitrary and non-literal interactions between prior knowledge in verbal form (which we call Subsumers[10]) and the ideas we meant to be teaching, turning our attention to the verbal content expressed by the students. We aimed at the variations on Subsumers, which were inferred from the Cognitive Structure variables changing and/or from the assimilation of new essential features by the Subsumers.

---

[10] There is no difference between Subsumers and explicit knowledge-in-action, once the second is verbalized.

The operatory knowledge analysis consisted in the search of possible knowledge-in-action used implicitly by the students when they were mastering different, but correlated situations. The inferences done from indicators, which are inherent to attempts of reconstruction of thought operations adopted by the students, were very important. Through this process, it was possible to understand these operations from lapses or from the apparently illogical using of symbolic operations (Pantoja, 2011). In the next section, we shall discuss the results of research.

**RESULTS**

We shall now present the results of this research. We divide this presentation in two sections. In one of them, which we call predicative knowledge analysis, we discuss the acquisition of predicative knowledge and in the other, which we call operatory knowledge; we discuss the mastering of situations using operatory knowledge.

**Predicative knowledge analysis**

For both construction and implementation of the didactic approach, we supposed some prior knowledge carried by the students since the beginning of the instruction. We took for granted that students can know the concepts of position, velocity, energy (specific Dynamical Variables), path (classical determination of the State of a Physical System); and systems like spring-mass, planetary, hydrogen atom (specific systems), Bohr's atom. These considerations have played an important role in the teaching process. We then started trying to facilitate conceptual super ordination from the prior knowledge of students.

In the pre-test, we mapped the criterial attributes used by the students when dealing with the concepts we were meant to be teaching. We describe these criterial attributes below, but we shall clarify first the way in which we labeled them. The letters associated to each attribute indicates the concept to which an attributes refers (**S** for Physical System, **V** for Dynamical Variable, **E** for State of a Physical System, **T** for Time Evolution). The numbers indicate different Meanings attached to each one.

We can see in the table below, that there was some regularity in the essential attributes showed by students in the pre-test. The level of generality of the concepts was essentially very high and this indicates general, abstract and, in this case, even poor concepts of Physical Systems, Dynamical Variables, State, and Time Evolution. Answers were considerably different, but expressed regular ideas.

*Table 1:* Level of generality of the students' concepts of Physical System, Dynamical Variable, State of a Physical System and Time Evolution in the pre-test.

| **Student** | *PS* | *DV* | *SPS* | *TE* | *Class.* |
|---|---|---|---|---|---|
| *Jerry* | 1.S | 1.E | | 1.T | *Very General* |
| *Layne* | | | | | |
| *Demri* | 2.S | 1.V | 1.E | 1.T | *Very General* |
| *Sean* | 2.S | | | 1.T | *Very General* |
| *Mike* | 2.S | 2.V | 1.E | 1.T | *Very General* |
| *William* | 2.S | 3.V | | | *Very General* |

We shall point what is the Meaning of the enumerated essential features of the concepts. Let us start by the concept of Physical System, for which we found two instances for its Meaning. For it we read: **1.S.** Part of the nature that is chosen for scientific inquiry; **2.S.** Theoretical Model that describes reality.

Comparing both attributes, we concluded the second one is more general[11] than the first, once it's not specified in **2.S** what they understood by theoretical model. In **1.S,** on the other hand, students stress that a Physical System is related to a domain or part of nature chosen for scientific inquiry. In summary, the first one was considered as highly general and the second one, more abstract than the first.

For the concept of Dynamical Variable, we enumerated two attributes, **1.V.** and **2.V.** The first one is described as "Physical variables that change in time" and the second one is understood as "Physical Properties that are correlated as time passes by". We considered the attribute **1.V.** the least general, because it goes directly to the core of the concept, which is the time variation of a physical quantity. The attribute **2.V.** emphasizes some correlations between variables but do not specify anything more than that. In sum, the first, although general, is the least abstract one.

For the concept of State of a Physical System, we enumerated just one attribute, namely, **1.E.** Its Meaning is associated to the configuration of the system in a specific instant. So, we tagged it "How the system is" and attached to it a very high level of generality, once this statement doesn't entail any example or extension of the idea of "being".

Finally, for the concept of Time Evolution, the most specific, epistemologically speaking once it is associated to the others, we found just one attribute we, **1.T** (Pantoja et al., 2012; Pantoja et al. 2014)**.** It's dubbed "Changes of the features that define the state" and considered it of a high level of generality, once it states the keystone idea of Time Evolution.

Every concept has a level of generality, but, in this case, the variation among students' concepts seems to be low. To do this classification we searched physical essential features, for example, relevant actions and subjects (as stressed for the attribute **1.T.** and for the attribute **1.V.**). The mean generality level considered the number of essential attributes in a concept: the more specific a concept, the greater its number of essential attributes are (Pantoja, 2011; Pantoja et al., 2012; Pantoja et al. 2014).

In this analysis, we reached the same conclusion of our prior work on Quantum Mechanics Teaching (Pantoja, 2011; Pantoja et al., 2012; Pantoja et al. 2014): the level of generality attached by the students to these concepts was very high. These evidences indicate few differentiation and great limitation for Meaningful Learning processes, once prior knowledge is barely relevant to the instruction. Since the instruction focused the teaching of elementary concepts, we understood that it was possible to present these students new and more differentiated criterial attributes.

we will discuss below the conceptual evolution of students' ideas throughout the instruction, triangulating the problem-solving activities with information in their concept maps. More detail on the concept maps analysis can be found on Pantoja (2011). This analysis embraces the study of acquisition and retention of essential attributes of the concepts involved in the proposal. We made a synthetic table in which are showed its essential attributes.

The criterion used for designing the table is relating essential attributes to concepts attached to the six tasks. As we know, concepts are tied one to another and can be basically remembered in two ways: induced (when the problem mentions a concept that needs to be remembered) or spontaneously (when students spontaneously remember a concept in order to answer a question for which he/she judges it is important). The main themes of the tasks, followed the sequence below:

- 1 - Physical System and Dynamical Variables;
- 2 - Dynamical Variables;
- 3 and 4 - State of a Physical System;

---

[11] The generality level is a general account for conceptualization in pre-test. Jerry, for example, used a highly general concept for Physical Systems, but for the other concepts the attributes were more general than the first mentioned. So the mean value is very high level.

- 5 and 6 - Time Evolution.

We also built a convention for reading the essential attributes, and it's a sign of three digits. The first digit represents the task in which the feature has been mapped: **a** - first task; **b** - second task; **c** - third task; **d** - fourth task; **e** - fifth task; **f** - sixth task. The second digit represents the class in which attribute is included. The last one represents the concept to which the attribute mentioned is linked: **S**. Physical System, **V**. Dynamical Variable, **VI (C)**. Incompatible (Compatible) Dynamical Variables, **E**. State of a Physical System, **SE**. State superposition; **T**. Time Evolution. This kind of classification follows the same logic of a prior work on the same topic (Pantoja et al., 2012). We shall now turn explicit the essential attributes mapped for the students.

The attributes (Pantoja et al., 2014) related to the concept of Physical System are: **a.1.S.** - Physical System is a set of objects that interact accordingly to specific laws; **a.2.S.** - Physical System is a physical phenomenon we are to describe; and **c.1.S.** which is equal to **a.1.S**.

The attributes related to the concept of Dynamical Variables are: **a.1.V.** Dynamical Variables are physical quantities; **b.1.V.** which is equal to **a.1.V.** and **b.2.V.** Measurement has the same Meaning of determination; **b.3.V.** Dynamical Variables are physical quantities that change as time passes by; **b.1.VI(C)** Incompatible (compatible) Dynamical Variables cannot (can) be measured simultaneously; **b.2.VI(C)** Incompatible (compatible) Dynamical Variables are not measurable in the same measurement process; **b.1.VI.** What characterizes the processes of measurement of Incompatible Dynamical Variables is the destruction of prior information of one of them when the measurement of the second is carried out after it is measured; **b.2.VI.** Incompatible Dynamical Variables satisfy the uncertainty principle; **b.3.VI.** Compatible Dynamical Variables do not commute; **b.X.VI.** The uncertainty principle includes information of variations and not of dispersions; **c.1.V.** That is equal to **b.3.V.**; **d.1.V.** The Hamiltonian is a Dynamical Variable identical to the energy; **d.2.V.** The Hamiltonian holds important information about interactions occurring in the system **e.1.V.** Classical Dynamical Variables are simultaneously determined; **f.1.V.** Incompatibility among Dynamical Variables propagates throughout time.

The features related to the concept of State of a Physical System are: **c.1.E.** State of a Physical System means a set of Dynamical Variables measured in a determinate instant; **c.2.E.** Eigenstates are associated to possible values of Dynamical Variables obtainable in a measurement process; **c.3.E.** State of a Physical System is the representation of the properties of a Physical System; **c.4.E.** State of a Physical System is a constrained set of Dynamical Variables that give us the maximum information possible; **c.5.E.** We cannot know the state, once it's probabilistic; **c.1.S.E.** Superposition of States implies the coexistence of two or more values of a given Dynamical Variable; **c.2.S.E.** State Superposition is the linear combination of two (or more) states **d.1.E.** The State of a quantum system is undetermined, because of its uncertainties; **d.2.E.** The Hamiltonian operator represents the State of a Physical System; **d.1.SE.** The state superposition is associated to the Incompatible Dynamical Variables, because when we determine one of the variables, the one becomes undetermined. **d.2.SE.** The state superposition is the occurrence of two simultaneous values of a Dynamical Variable; **e.1.E.** The Physical State of a System is a superfluous entity: it resumes itself to a set of Dynamical Variables; **e.2.E.** The state is a time-varying description of a Physical System.

The features related to the concept of Time Evolution are: **e.1.T.** The Physical System is the entity that suffers the Time Evolution; **e.2.T.** Time Evolution has interactions as physical cause; **e.3.T.** Time Evolution in Quantum Mechanics and in Classical Mechanics depends on initial conditions. In Quantum Mechanics this is translated in terms of probabilities; **f.1.V.** Incompatibility among Dynamical Variables propagates throughout time; **f.1.T.** Schrödinger's equation describes Time Evolution of the state, keeping stationary the Dynamical Variables. The table below summarizes the essential features of the concepts we wished to teach.

***Table 2:*** Essential features attached to the students to the concepts of Physical System, Dynamical Variables, State of a Physical System and Time Evolution during time

| *Student* | *Physical System* | *Dynamical Variables* | *State of a Physical System* | *Time Evolution* |
|---|---|---|---|---|
|  |  |  |  |  |
| *Jerry* | a.1.S, c.1.s. | a.1. V, b.1.V, b.2V, b.1.VI(C)., b.1.VI., b.2.VI., c.1.V, d.1.V, d.2.V. | c.1. SE, c.1.E, c.2.E, d.1.SE | - |
| *Layne* | a.2.S, c.1.S. | a.1. V, b.2.V, b.1.VI., b.2.VI.,b.3.VI, d.2.V, e.1.V. | c.1. SE, c.4.SE, d.2.SE, e.2.E | e. 1.T |
| *Demri* | a.1.S, c.1.s. | a.1.V, c.1.V | c.1.E | - |
| *Sean* | a.1.s. | a.1.V, b.1.V, b.3.V, b.1.VI(C), b.X.VI, b.2.V, d.1.V, d.2.V | c.1.E, c.5.E, d.1.E, e.1.E | e.2.T, e.1.T |
| *Mike* | a.1.S, c.1.S | a.1. V, b.2.VI(C), b.1.VI, b.2.VI, b.2.V, c.1.V, d.2.V,f.1.V | e. 3.T, f.1.T | e. 3.T, f.1.T |
| *William* | a.1.S, c.1.S | a.1. V, b.3.VI(C), b.2.V, c.1.V, d.1.V | c.3. E | - |

From now on, we will discuss the acquisition of each concept in a global form, in other words, we will point similarities and differences between essential attributes of the concepts as mastered by the students. A more local approach can be found in Pantoja (2011), a work in which System → Dynamical Variable → State of a Physical System → Time Evolution.

*Physical System*

We can see a strong uniformity in the use of the attribute a.1.S., which carries the Meaning of being a set of objects in which components act upon each other according specific laws. This view is coherent with the one taught in classroom.

"*Systems of fluids: they have internal structure constituted by atoms and molecules. They present Dynamical Variables as pressure, temperature and internal energy…*" (Jerry)

"*The Earth and the Sun system; Objects: Earth and sun; interaction: gravitational; it's important for the comprehension of the movement of celestial bodies.*" (Sean)

Just one of the six students (Layne) confused the concept with the one of Physical Phenomenon (feature a.2.S.). Layne doesn't mention this with all the words, but exemplifies the concept of Physical System mentioning physical phenomena and this is evidence that it is a possible Theorem-in-action of which he is conscious, although it is almost implicit for him, that we will discuss in more detail in the next section.

"[Give one example of Physical Systems] *A collision between two small spheres. In this collision, there is internal structure, but it's not significant to our interest, which is the collision…*" (Layne)

In the third task, we could note some stability of the concept of Physical System. This may indicate that students' prior knowledge, that contained specific examples of Physical Systems, was stable enough to superordinate the more general concept of Physical System. One can note, from the examples given by the students, that the presentation of situations belonging to different areas of Physics seemed to facilitate the acquisition of concepts.

Another important point to mention is the generality level of Quantum Physical Systems conceptualization is higher than the one of the classical concept's counterpart. This is discussed in more detail in Pantoja (2011).

*Dynamical Variables*

In a prior work we observed relative agreement in students' conceptualization on the concept of Dynamical Variables (Pantoja et al., 2012). For this one we found a similar result. The conceptual super ordination has been accomplished, because the students may have assimilated the main feature of the concept of Dynamical Variables, strictly speaking, the one of being time-varying physical quantities (a.1.V.). We must stress that the students did not emphasized the time-varying aspect in the first task, but half of them did it in the third one. We infer, from these findings, that students considered this attribute obvious, so they use it implicitly. As we will discuss in the operatory knowledge analysis, Time appears as a Concept-in-action and it's coherent with this point.

"*A System of fluids: they have internal structure made of atoms, molecules. They also are associated to Dynamical Variables as, for example, pressure, temperature and internal energy*" (Jerry)

"*Main Dynamical Variables: velocity, position, potential and mass*" (Mike)

It's notable the differentiation of explicit students' concepts. In the second task, we found just one shared attribute. Three students related the Incompatibility among Dynamical Variables to destruction of prior information obtained of one of them in a process of sequential measurement (b.1.VI.), for example, $x$ and $p_x$ measurements or $s_x$ and $s_y$ measurements. Another point to stress is the hypothesis on the obviousness of the time-varying feature raised above. Few students have explicitly emphasized this aspect. We see, in the other hand, that four students "remembered" the time-varying feature in the third task. From all these cues, we infer that it is not just remembering that is occurring in this process, but the implicit consideration of Dynamical Variables as time-varying entities.

"*... it's like we have an 'uncertainty' related to the measurement of the spins in certain directions.*" (Layne[12])

"*By measuring one of the components, we change the other ones*"(Mike)

Another feature that should be discussed is the consideration of the Hamiltonian as a Dynamical Variable identical to the Energy (d.1.V.), an association made by three students. In first place, it represents an extended combination of the concepts of Dynamical Variable and Energy to the concept of Hamiltonian, which was essentially new to the students. Despite being an erroneous association, this might be evidence of Meaningful Learning. On the other hand, we must be cautious with this kind of conception that must be negotiated.

> "*The hamiltonian operator is used in Quantum Mechanics and its compatible just with some Dynamical Variables. The hamiltonian function is used in Classical Mechanics and it's compatible with all Dynamical Variables*" (William)

The last feature we want to point out, because it's relatively shared by the students, is the one that considers the Hamiltonian gives information on the interactions occurring in the Physical System (d.2.V.). Although some students treat the Hamiltonian as a Dynamical Variable, we must acknowledge that they recognize the important information concerning interactions in systems. For conservative systems, the recognition is direct: the students relate the Hamiltonian to the interaction by means of the potential energy. We shall argue that it's an evidence of Meaningful Learning, because they base in their prior knowledge on potential energy to understand initially the Hamiltonian.

"*[Given the Hamiltonian associated to the Dynamics of two planets orbiting around a common center of mass]... the expression $\frac{GMm}{|r-R|}$" is the one associated to the interaction*" (Sean)

"*[Given a Hamiltonian that considers the interaction between a particle with magnetic moment and an inhomogeneous magnetic field] the particle interacts with the Magnetic Field*" (Sean)

---

[12] Fragment of the question: [... when we measure one of the Dynamical Variables, we lose prior information due to the other two measured before. How do you understand this statement?]

We shall now discuss some specific but noticeable problems in understanding. The feature b.X.VI reveals some confusion in conceiving the uncertainty principle. One student (Sean) treats it as a classical uncertainty, since he associates the uncertainties in momenta, as accelerations. Other point to enlighten is the attribute b.2.V. that is confusion between measurement and determination. Students often refer to determination of Dynamical Variables as measurement. This difference has been stressed, but somehow the students couldn't understand the difference between arranging and executing a measurement (to measure) and being in an eigenstate of a given operator, in other words, knowing a specific value for a Dynamical Variable (to determine). In a prior study, we found this students' conception more frequently (Pantoja et al., 2012).

*"... you can only know the position of an object if it is still; otherwise, it would interfere in the velocity"* (Sean)

*"Compatible Dynamical Variables are the ones that can be **measured** simultaneously... the process of measurement of one of them don't alter the other one"* (Jerry – our emphasis)

*State of a Physical System*

The students seem to share a minor set of Meanings related to the concept of State of a Physical System. In prior work, students started to acquire essentially different Meanings when the discussion reached the concept of State of a Physical System, an intrinsically abstract and unknown concept for them. In Classical Physics, it is treated implicitly and that's a possible reason for students to develop very idiosyncratic conceptions on this concept. Despite this conceptual division, we will discuss some points that are noticeable.

Most of the students (five out of six) anchored the concept of state as an elaboration of the one of Dynamical Variables (c.1.E.). This is partially important, because we can evidence a pattern of Meaningful Learning (learning from something relevant they already know), but in terms of content, students tend to conceive State of Physical System as a set of Dynamical Variables and not as a configuration of these variables. Instead of numbering sets of Compatible Dynamical Variables and determining the probabilities amplitudes for occurrence of certain values for these Variables during a determination process, they tend to understand that the state as equal to the Physical Quantities. We can see this feature as the more general of the students' conceptualization.

*"In Quantum Mechanics, the Dynamical Variables are not always measured simultaneously"* (William[13])

The idea of State Superposition as a coexistence of two values is very common when people intent to interpret Quantum Mechanics and discuss the Schrödinger's cat thought experiment (c.1.SE.). We cannot say that this is indeed incorrect, but we can say it is an interpretation we did not emphasize on the phenomena. A hypothesis that can be sustained is that they meant to say that the value of a certain Dynamical Variable to be measured is not actually defined when the state of a System is written as a Superposition of Eigenstates of the Operator relative to this variable.

> *"By state superposition, I understand a system in which we have two different kinds of states. Its' physical interpretation concerns to the fact that a Physical System can assume different values for a certain Dynamical Variable"* (Layne)

The attribute that relates the two specific quantum concepts of Incompatible Dynamical Variables and State Superposition (d.2.SE.) is of great importance, once it relates the aspect of uncertainty, in a relatively adequate way, to the Dynamical Variables and to the State Superposition principle. Two students expressed it in an explicit form and this relation must, obviously, be fostered and differentiated to elaborate the notion of Incompatible Dynamical Variables and of Uncertainty Principle.

*Time Evolution*

---

[13] *Fragment of question: [How the Meaning of the concept of state is different in Classical Mechanics and Quantum Mechanics?]*

We shall now stress that three students (Jerry, Layne and Sean) didn't answer to the sixth task and one (Jerry) didn't answer to the fifth task. We recognize that mortality occurred in these cases. This damaged our conclusions on the analysis of the last two tasks in the predicative form. Nevertheless, we could enumerate some conceptions regarding this aim-concept.

Two students shared the idea that Time Evolution is attached to Physical Systems themselves (e.1.T.). This can be seen in two different ways: 1) Linguistic economy, since saying "Time Evolution of the State of a Physical System" is longer than doing the same with "Time Evolution of the Physical System", keeping implicit the fact that the Time Evolution is attached to the system; 2) The consideration of the State as a superfluous entity, because it's composed of Dynamical Variables and, so, the Time Evolution feature passes to the System, once the Dynamical Variables describe them. Both hypotheses can be sustained, but we shall discuss if one of them is more probable for each student. These two students are Layne and Sean.

We now trace back to their characterization of State of a Physical System. Sean conceives the state as being an unknowable entity, because of the quantum uncertainties (d.1.E.) and probabilities (c.5.E.). He also believes that the state is a set of Dynamical Variables attached to the system (c.1.E.). For the state cannot be known, but the Dynamical Variables do, and the latter describe the system, we tend to conclude that Sean attributes the Time Evolution to the System itself.

Layne conceptualizes in a slightly different way. He conceives that the state is restrained to the knowledge of the definite values of some Compatible Dynamical Variables and this is the maximum information that we can extract from the system (c.4.E.). Besides this feature, he attributes the time-varying characteristic to the state when conceptualizes it as a time-varying description of the System. So, for Layne's case we can argue that he used linguistic economy to refer to the state.

An interesting conception evidenced in this study is the one emphasizing the role of initial conditions in the determination of the State, both in Classical and Quantum Mechanics (e.3.T.). We know that, unless a measurement occurs, the state evolves in time in a causal way. Mike was the only student that evidenced comprehension regarding this feature[14].

*"... In Quantum Mechanics the equation of State as a function of time depends on the initial state"* (Mike)

We shall now discuss the possible knowledge-in-action carried implicitly by the students in the problem-solving processes. It's interesting to establish a connection between operatory and predicative forms of knowledge. As we shall see, students can solve problems despite their difficulties in dealing with predicative knowledge in Quantum Mechanics. Some of their theorems-in-action provide enlightening in the problem-solving processes while other makes it harder.

**Operatory Knowledge Analysis**

When we discuss operatory knowledge, it's unconceivable to separate the categories Vergnaud calls Concept-in-action from the theorems-in-action. This can be understood both as a dialectic character of categories and propositions, and of interconnection between different categories through the mediation of propositions. The latter point is stressed because we want to emphasize the interaction between situations (tasks that are problems) and schemas (stable cognitive mechanism for initially mastering of situations).

In the predicative form, although it was very clear the concepts were tied together, we could think on a relationship between an abstract object and Relevant Prior **Knowledge** of the student. In other words, a student could refer to each category (Time-Evolution, Physical System, Dynamical Variable, State of a Physical System) independently of the context of the task. We could ask them, "what means the state superposition principle?", for example. They would refer directly to this object, in an abstract form, which characterizes the subject-object interaction stressed by Piaget. That's the main difference between formalized and non-formalized knowledge.

Vergnaud states that a great number of psychologists have cast aside or even despised the Operatory Knowledge and the first stage of conceptualization, the interaction between Schemas and

---

[14] Obviously, someone could accuse us of a huge mistake, but the context of the question involved no measurements along time. Once a measurement is done, Quantum Mechanics ceases to be strictly causal (Messiah, 1999).

Situations. To consider this latter form of knowledge is a difficult task, because operations occurring are implicit and so it's necessary to infer which knowledge-in-action are carried by a Schema. We show in the table 3 below the possible Knowledge-in-action carried by the students. We now dub these Theorems-in-action (that include Concepts-in-action marked in boldface) by: 1) a letter indicating the student that carries the referred knowledge-in-action; 2) a number that indicates the knowledge-in-action we refer. For example, the second Theorem-in-action used by William is E.2. We couldn't evidence any knowledge-in-action for Demri, because she just copied and pasted phrases from the instruction material we built and used in the proposal[15]. Jerry is identified as A, Layne as B, Sean as C, Mike as D and William as E.

We shall now discuss some shared theorems-in-action of the students. There are two theorems-in-action shared by two students. They're dubbed A.1 (B.2) shared by Jerry and Layne, and C.3 (E.2) shared by Mike and William. We shall discuss other theorems-in-action, in special the ones of William, the student that had the lowest level of predicative knowledge. Three out of four possible theorems-in-action are used in a wrong or incomplete way. There are two other possible theorems-in-action, one of them due to Mike and the other one due to Sean, that also must be discussed. The Theorem-in-action A.1., stated[16] by Jerry and Layne, is coherent, however it's general. This possible Theorem-in-action links the sequential measurement processes in the Stern-Gerlach experiment to the Incompatible Dynamical Variables, by the splitting of a beam of silver atoms. Therefore, it was essential in the problem-solving processes. It must be stressed that this is strongly related to the understanding of the uncertainty relations for the spin projections onto $x$, $y$ and $z$.

The Theorem-in-action C.3. stated by Sean carries the physical concept of probabilistic Quantum Measurements. We point that it must be enhanced by presenting more situations for the students to both make it explicit and differentiate its' Meaning. Probability has central role in Quantum Measurements, but it must be understood as more than simple probabilities associated to stochastic phenomena, in other words, knowledge constructed to understand nature is probabilistic, but probability in Quantum Mechanics doesn't mean ignorance about interactions we don't know how to exactly describe, but something on which Quantum Theory is constructed. One must be careful in this differentiation, once probabilities do not invalidate our possibility of foreseeing the results of a given measurement. Sometimes and in certain configurations (preparation of the state of the system in stationary energy eigenstates, for example) the predicted result is attached to probability equal 1, and even when we don't have it, we can foresee probabilities for the results of the measurements.

Another Theorem-in-action, used by William, is the one dubbed E.4 which states that "a quantum object follows a definite path". To attribute a definite path to quantum objects is equivalent to deny the position-momentum uncertainty principle and this was stressed throughout the didactic approach. We can understand this situation in diverse complementary ways: 1) the stability of the classical ideas in the Cognitive Structure of the student and use of mechanistic ideas (Mashadi, 1996 apud Greca and Moreira, 2001), 2) her tendency to rote learning[17], 3) not pointing the differences between Classical Physics and Quantum Physics (Gil and Solbes, 1993, apud Greca and Moreira, 2001). We must stress that this result was previously obtained by Nidderer et al. (1990).

The Theorem-in-action D.4, stated by Mike, is the one we will discuss a little more. Its' physical content expresses that electrons, which are quantum objects, can be understood through Classical Electrodynamics. In the classes, he has argued that we could understand electrons as hard spheres charged with the elemental electric charge and, in some problems, he mentioned the calculation of simultaneous velocity and position. We infer that it's an implicit Theorem-in-action that we are not sure that can get explicit. This student was aware of the Incompatibility between Position $x$ component and Momentum $x$ component, but he was used to speak in Classical Electrodynamics language when he figured out the situation for quantum electrons. This can be attached to a mechanistic thinking as stressed by Mashadi (1996).

The Theorem-in-action C.1., stated by Sean, states that if a beam of silver atoms whose state indicates that a measurement of the $s_z$ spin component returns two beams ($s_z,+$ and $s_z,-$) of given rate of intensities, the sequential measurement of $s_x$ or $s_y$ will return two beams of the same ratio in intensities. For example, if we imagine the first split as being composed of a beam with $I(s_z,+) = 0{,}8I_0$ and $I(s_z,-) = 0{,}2I_0$, following the Theorem-in-action of Sean, a sequential measurement of $s_x$ of the beam $s_z,+$ will return a beam $s_x,+$ with intensity $I(s_x,+)(s_z,+) = 0{,}64I_0$ and a beam $s_x,-$ with intensity $I(s_x,-)(s_z,+) = 0{,}16I_0$. Now we follow to the discussion and pedagogical implications.

---

[15] The researchers wondered if this could be even considered as rote learning. There was a consensus in not classifying it as a learning process.
[16] By "state" we mean to use implicitly a given idea. That's why we characterized it as an implicit Theorem-in-action.
[17] Expressed in her random answers.

*Table 3:* Theorems-in-action used by students in the mastering of Situations

| | | | | | |
|---|---|---|---|---|---|
| A | A.1. In the Stern-Gerlach experiment, every time we measure **Incompatible Dynamical Variables**, there is a splitting in the beam of silver atoms | | | A.2 **Quantum systems** are random. | A.3. **Measurement** makes the state of the Physical System to jump in one of the random **eigenstates**. |
| B | B.1. In the **Measurement** process, we obtain one of the possible values for a **Dynamical Variable** which is measured. | B.2. = A.1. | B.3. **Physical Systems** are Physical Phenomena | B.4. The **state of the system** informs the **intensity** of the resulting beam in the Stern-Gerlach experiment | B.5. **Eigenstates** of an operator associated to a **Dynamical Variable** are mutually exclusive. |
| C | C.1. The **intensities** of the beams resulting of a second **measurement** in a Stern-Gerlach experiment are proportional to the **intensities** resulting of the first **measurement**. | | C.2. The **measurement** of a **Dynamical Variable** alters the value of the first measured. | C.3. In Quantum Mechanics, **measurement** is a **probabilistic** process. | C.4. Interactions are described by fundamental equations and Time Evolution is caused by the former |
| D | D.1. Simultaneous **measurements** of variables occur, literally, at the same time and not sequentially. | | D.2. The **state of the Physical System** is modified throughout time. It's described by the Schrödinger's equation; whose solution depends on initial conditions. | D.3 Projection Postulate. | D.4 Electrons can be understood completely in Classical Electrodynamics. |
| E | E.1. **Quantum objects** are **non-localizable** | E.2. = C.3. | E.3. When a beam of silver atoms passes through a Stern-Gerlach apparatus, the **spin** is divided in equal **intensities**. | | E.4. **Quantum objects** follow a definite **path** |

**DISCUSSION**

There was strong uniformity in the mastering of the concept of Physical System, although its' quantum form seems more abstract to students than its' classical counterpart. One subject (Layne) initially showed confusion between Physical System and Physical Phenomena. Layne, however, achieved another level of conceptualization with respect of this concept.

For the concept of Dynamical Variable, the same uniformity was just evidenced for the feature of being physical quantities, something very related to their prior knowledge, what shows us evidences of conceptual super-ordination. The students could understand also the essential difference compatible and incompatible Dynamical Variables.

With respect to the concept of incompatible Dynamical Variables, the prior information destruction feature played important role in the differentiation between compatible and incompatible Dynamical Variables. Students tend, however, to identify measurements and determinations. In other words, they conceive the expressions have the same Meaning.

The consideration of students on the role of the Hamiltonian as describing interactions occurring in a Physical System must be stressed if possible, but it is a common association students may establish that the Hamiltonian operator is as a quantum Dynamical Variable identified to the energy. This can be easily discussed with situations in which the Hamiltonian Operator includes time dependent interactions.

Heterogeneity on conceptualization on the concept of State of a Physical System must be discussed. The concept is not attached directly to the students' relevant prior knowledge and it's very abstract in its nature. Considering it is not included in Classical Mechanics discussions in the context of the students we worked with, we inferred that the huge heterogeneity is associated to the attempt of making sense of this very new concept.

There are more positive points to stress. There is evidence that the students followed Meaningful Learning patterns, once the concept of State was understood as an elaboration of the concept of Dynamical Variables. It must be pointed that more emphasis on the differentiation between these two instances must be done, because if two concepts are regarded as equal, one of them shall obliterate (Ausubel, 2000), and it seems that some students, in some cases, meaningfully learned the concept of State of a Physical System.

Some students consider that State Superposition a coexistence of two values of a given Dynamical Variable, which is not always true, since, for example, we may have a $|s_x,+\rangle$ state, with $s_x$ determined, which can be understood as $|s_x\rangle = \frac{1}{\sqrt{2}}|s_z,+\rangle + \frac{1}{\sqrt{2}}|s_z,-\rangle$. But regarding we discussed the concepts briefly, conceiving things this way can be considered a good achievement.

It must be stressed that most of the students explicitly associated time to Dynamical Variables, but did not ever mentioned the time-varying aspect of the State (except Layne and Mike). It must be pointed too that some students attach the Time Evolution to the Physical System itself for two possible reasons: 1) linguistic economy or 2) consideration of the state as a superfluous entity.

Another important feature grasped just by one student in predicative knowledge is the dependence of the State on initial conditions. This is an important point and it's closely related to prediction in Quantum Mechanics.

We can also see that operatory knowledge play an important role in Quantum Mechanics. Besides that, if it's not turned explicit by language, in other words, made predicative, it can't be discussed (Vergnaud, 1983), and some of them may operate as hindrances for learning.

An example of theorems-in-action that hinder learning is the Theorem-in-action stating that Quantum Measurements are always probabilistic. This can be an obstacle for the learning of stationary energy eigenstates. An example of Theorem-in-action fostering learning is the one that entails the splitting of the beams of Stern-Gerlach experiment in the sequential measurement of incompatible Dynamical Variables, stated by Jerry and Layne.

We must stress that few Theorems-in-action were shared by the students. This is a feature that reveals the private characteristic of these implicit constructs and the importance of turning them explicit.

Based on the conclusions of this and of correlated studies (Pantoja et al., 2012a; Pantoja et al., 2012 b; Pantoja, 2011) we suggest some pedagogical implications for Quantum Mechanics teaching:

- Presenting potentially Meaningful situations is likely to facilitate both construction of fruitful Theorems-in-action and conceptual super-ordination in Quantum Mechanics, by discussing similarities and differences between Classical and Quantum Mechanics;

- Progressive differentiation and integrative reconciliation are fruitful principles to be followed in instructions in Physics Teaching. To stress both ruptures and threats may lead students to perceive the frontier between Classical Physics and Quantum Physics;

- Prior knowledge, as Ausubel stresses, is something that must be seriously taken into account. Some students have Classical Relevant Prior Knowledge and extending the discussion on Physical Quantum Systems must be an important point to regard. It's unavoidable to take some things for granted with respect to students' prior knowledge. But clearness in the concept of Quantum Physical System is not something we can consider, and sometimes is useful to present a great variety of situations of Quantum Physics to facilitate adaptation of the students in this conceptual field;

- Some theorems-in-action and concepts-in-action constructed are hindrances for learning while others serve as good implicit strategies for problem-solving. It's likely that operatory knowledge influences upon predicative knowledge acquisition, once the former seems to be the first source of conceptualization. It's important to study how this influence is made. It's important to turn these entities explicit, because as Vergnaud stresses, just explicit propositions are debatable;

- Predicative knowledge and operatory knowledge are important in Quantum Mechanics teaching. The first is the source of Meaning negotiation, but composes the last level of verbal conceptualization. It's fruitful to work on its acquisition from various aspects. We emphasized problem-solving and Meaning negotiation, but other processes like modelling activities, collaborative activities or oriented-inquiry teaching may also help the process. It is necessary to take into account that operatory knowledge, however, composes most part of cognition, once it is merged with actions.